\documentclass[prd,twocolumn,amsfonts,showpacs]{revtex4-1}

\pdfoutput=1

\usepackage{epsfig}
\usepackage{psfrag}
\usepackage{amsmath}
\usepackage{amssymb}
\usepackage{color}
\usepackage{hyperref}

\begin{document}
\author{Wei-Can Yang$^{1}$}
\author{Chuan-Yin Xia$^{2,1}$}
\author{Hua-Bi Zeng$^{1}$}
\email{hbzeng@yzu.edu.cn}
\author{Hai-Qing Zhang $^{3}$}

\affiliation{$^1$ Center for Gravitation and Cosmology, College of Physical Science
and Technology, Yangzhou University, Yangzhou 225009, China}
\affiliation{$^2$ School of Science, Kunming University of Science and Technology, Kunming 650500, China}
\affiliation{$^3$ Center for Gravitational Physics, Department of Space Science \&
 International Research Institute for Multidisciplinary Science, Beihang University, Beijing 100191, China}

\title{Phase Separation  and  Exotic Vortex Phases in a Two-Species   Holographic Superfluid}

\begin{abstract}
At a finite temperature, the stable equilibrium states of a  coupled two-component superfluid
with the same mass in both non-rotating and rotating cases can be obtained by studying its real time dynamics via holography,
the equilibrium state is  the final stable state that does not change in time anymore in the evolution process .
Without rotation, the spatial phase separated states of the two components
become more stable than the miscible condensates state when the direct repulsive inter-component coupling constant $\eta>\eta_c=0.05$
when the Josephson coupling $\epsilon$ is turned off. While a finite $\epsilon$ will
always prevent the two species to be separated spatially.
Under rotation, with vanishing $\epsilon$, the quantum fluid reveals many equilibrium structures of vortex
states by varying the $\eta$ from negative to positive, the interlaced vortex
lattices undergo a phase transition to vortex sheets with each component made up
of chains of single quantized vortices.
\end{abstract}
\maketitle
\section{Introduction}
The gauge-gravity duality\cite{Maldacena,Witten,Gubser} that relates
strongly interacting quantum field  theories to theories of classic gravity
in higher dimensions has provided a new scheme not only to study
 strongly interacting condensed matter systems in equilibrium\cite{Zaanen:2015oix,Ammon2015},
but also to study the real time dynamics when the system is far away from equilibrium \cite{Liu:2018crr,Liu,Sonner:2014tca,Bhaseen,Chesler:2014gya}.
The first proposed theory of a single component holographic superfluid/BEC was given in \cite{Gubser2008,Hartnoll,Herzog}.
The array of vortices is known to happen in a
rotating superfluid\cite{Feynman}.
To find a static vortex lattice solution in the presence of rotation without studying the dynamic process in holography is always very hard technically, an  only single vortex has been  obtained in the single component
system in \cite{Montull,Domenech,Keranen,Maeda,Dias,Tallarita}. However,
by studying the full dynamics of the single component holographic superfluid/BEC in a  rotating disk, the equilibrium vortex lattice can be obtained as the final time independent solutions in the real time dynamics process \cite{Xia:2019eje,Tianyu},
Besides the superfluid/BEC with only one order parameter,
a two-component superfluids/BECs with two coupled order parameters has also become one of the most concerned topics in condensed matter physics,
since it demonstrates novel quantum states that can not be observed in a single component system as predicted in the frame work of two component Gross-Pitaevskii (G-P) equations\cite{Kasamatsu-review}. The two components $\Psi_1$ and $\Psi_2$ are coupled through a direct coupling term $\eta |\Psi_1|^2|\Psi_2|^2$
and a Josephson coupling $\epsilon(\Psi_1\Psi_2^*+\Psi_1^*\Psi_2)$.
Without rotation, a two-component BECs  will enter a spatial separation
state for the two order parameters due to the repulsive interaction $\eta>0$ between the two components\cite{Shenoy,Bohn,Timmermans,Cornell}.
In the presence of rotation, by solving the time dependent two-component G-P equation, the equilibrium vortex structures can be
 obtained after a sufficient long time revolution, which reveal rich  structures  by varying the direct coupling  between the two components. As $\eta$ increase the interlocked vortex lattices undergo phase transitions from triangular to square, to double-core lattices, and eventually
develop  vortex sheets\cite{Ueda,Kasamatsu}. Vortex sheet is a state  that  the vortices are aligned to make up winding chains of single quantized vortices, and the chains of two components are interwoven alternately, or form as the cylindrical vortex sheets.
Such a solution was proposed for the first time by Landau and Lifshitz \cite{Landau} and has been observed experimentally in $^{3}$He A\cite{Parts}.
Vortex sheet has proved to be an important physical object with nontrivial topology, which has many physical applications\cite{Volovik}.

The purpose of this article is to research the equilibrium state properties of a strongly coupled two
component BECs both in the presence of rotation and without rotation by using the holographic method.
The static equilibrium states are calculated from the time dependent evolution equations,
where the static solution can be obtained by evolving the system from an initial homogeneous superfluid state
in the presence of a perturbation induced by rotation or a small fluctuation. Our treatment is very
similar to the calculation in the frame of a time dependent two component G-P equation,
where the static solution was found as the final state that does not change in time anymore\cite{Ueda,Kasamatsu}.
When there is no rotation, the phase separation states was studied in detail
for different temperatures and different values of coupling constant.
We found that the phase separation states prefer to appear at lower temperature
with a relatively higher inter-components direct repulsive interaction, while the
Josephson coupling is always prevent the system to be phase separated.
Under rotation,  the holographic
superfluid reveals four structures of vortex states corresponding different values of direct coupling between the
two components by turning off the Josephson coupling.
The four structures include a triangle lattice, a square lattice, a vortex stripe and a vortex sheet,
the vortex sheet solution is a result of phase separation.
The vortex diagram in the
intercomponent direct coupling $\eta$ versus rotation-frequency $\Omega$ is also calculated, which is  similar to the
 one obtained from the two component (G-P) equations \cite{Kasamatsu}.
Because of the nonlinear nature
of the gravity theory, we will have to rely on numerical
methods by solving a highly nonlinear partial differential gravity equations.

This paper is organized as follows. Section~2 introduces the holographic $U(1)$ symmetry broken model with two order parameters, the equations of motion (EoMs) for the model and also the numerical method we adapted to solve the EoMs. Section~3 discusses the properties of the  two component system
without rotation and  the phase separation state is obtained with a large repulsive inter-components coupling. Section~4 presents the
appearance of exotic vortex phases of the two-component system under rotation, and the vortex phase diagram is also obtained.
Section~5 is devoted to the conclusion and some discussions.

\section{Holographic model and the time evolution equations of motion}
The holographic model
we use is a bottom-up construction containing  two coupled charged scalar fields living
in an $AdS$ black hole background, which is
an direct extension of the single component holographic superfluid
model proposed in \cite{Hartnoll}. The action includes two charged scalar fields that coupled to a $U(1)$ gauge field\cite{Wen:2013ufa}
\begin{equation}
S=\int  d^4x \sqrt{-g}\Big[-\frac{1}{4}F^2-\sum_{j=1}^2( |D\Psi_j|^2-m_j^2|\Psi_j|^2)+ V(\Psi_1,\Psi_2)\Big],
\label{model}
\end{equation}
the inter-component coupling potential between the two charged scalar fields takes the form
\begin{equation}
V(\Psi_1,\Psi_2)= \epsilon(\Psi_1\Psi_2^*+\Psi_1^*\Psi_2)+\eta |\Psi_1|^2|\Psi_2|^2,
\label{potential}
\end{equation}
where $\eta$ is called the direct coupling while $\epsilon$ is the Josephson coupling.
Note that the interaction potential takes the same form as the G-P equations for a coupled two component superfluid  \cite{Ueda,Galteland,Nitta}. In the holography, there is another more complex  model dual to a two component superfluid with two gauge fields coupled to two charged scalar fields respectively \cite{ Bigazzi:2011ak,Musso}.

Also in the action, $F_{\mu\nu}=\partial_\mu A_\nu-\partial_\nu A_\mu$, $D_\mu=\partial_\mu-iqA_\mu$ with $q$ the charge. The metric of the $AdS_4$ black hole background in the Eddington-Finkelstein coordinates reads
\begin{equation}
ds^2=\frac{\ell^2}{z^2}\left(-f(z)dt^2-2dtdz + dr^2+ r^2d\theta^2\right).
\label{metric}
\end{equation}
in which $\ell$ is the AdS radius, $z$ is the AdS radial coordinate of the bulk
and $f(z)=1-(z/z_h)^3$.  Thus, $z=0$ is the AdS boundary while $z=z_h$ is the horizon; the Hawking temperature is $T=3/(4\pi z_h)$. $r$ and $\theta$ are respectively the radial and angular coordinates of the dual $2+1$ dimensional boundary, which is a disk that suitable to study the properties of the two component superfluid under rotation.

The axial gauge $A_z=0$ is adopted as in \cite{Liu,Herzog}.  Near the boundary $z=0$, by choosing that $m_1^2=m_2^2=-2$, the general solutions take the asymptotic form as,
\begin{eqnarray}
A_\nu &=& a_\nu + b_\nu z+\mathcal{O}(z^2),
\label{aboundary} \\
 \Psi_j&=& \Psi_j^1 z+ \Psi_j^2 z^2+\mathcal{O}(z^3).
 \label{psiboundary}
\end{eqnarray}
$j=1,2$, the coefficients $a_{r,\theta}$ can be regarded as the superfluid velocity along $r, \theta$ directions while $b_{r,\theta}$ as the conjugate currents \cite{Montull}. Coefficients $a_t$ and $b_t$ areinterpreted as chemical potential $\mu$ and charge density $\rho$  respectively in the boundary field theory. Moreover, $\Psi_j^1$ is a source term which is set to be zero, then $\Psi_j^2$ is the vacuum expectation value $\langle O_j\rangle$ of the dual scalar operator in the boundary in the spontaneous symmetry broken phase.
Without loss of generality we rescale $\ell = z_h = q=1$.
Therefore, by scaling $\Psi_j= \psi_j z$ and using the axial gauge that $A_z=0$, the equations of motion (EoMs) can be written as
\begin{widetext}
 \begin{equation}
 \begin{split}
 \partial_t\partial_z \psi_1 =\frac{1}{2}\{-\epsilon \frac{\psi_2}{z^2}-\eta \psi_1 |\psi_2|^2+i[\psi_1\partial_z A_t+2A_t \partial_z \psi_1-\psi_1 \partial_r A_r+2A_r \partial_r \psi_1-\frac{\psi_1\partial_\theta A_\theta +2A_\theta \partial_\theta\psi_1}{r^2}]\\
 +[(1-z^3)\partial_z^2 -3z^2\partial_z +\partial_r^2 +\frac{\partial_\theta^2}{r^2}-z-A_r^2-\frac{A_\theta^2}{r^2}]\psi_1\},
 \end{split}
 \end{equation}
 \begin{equation}
 \begin{split}
 \partial_t\partial_z \psi_2 =\frac{1}{2}\{-\epsilon \frac{\psi_1}{z^2}-\eta \psi_2 |\psi_1|^2+i[\psi_2\partial_z A_t+2A_t \partial_z \psi_2-\psi_2 \partial_r A_r+2A_r \partial_r \psi_2-\frac{\psi_2\partial_\theta A_\theta +2A_\theta \partial_\theta\psi_2}{r^2}]\\
 +[(1-z^3)\partial_z^2 -3z^2\partial_z +\partial_r^2 +\frac{\partial_\theta^2}{r^2}-z-A_r^2-\frac{A_\theta^2}{r^2}]\psi_2\},
 \end{split}
 \end{equation}
 \begin{equation}
 \partial_z^2 A_t =-2\Im(\psi_1^*\partial_z\psi_1+\psi_2^*\partial_z\psi_2)+\partial_z\partial_r A_r+\frac{\partial_z\partial_\theta A_\theta}{r^2},
 \end{equation}
 \begin{equation}
 \partial_t\partial_z A_r=\Im(\psi_1^*\partial_r\psi_1+\psi_2^*\partial_r\psi_2)-A_r(|\psi_1|^2+|\psi_2|^2)
 +\frac{\partial_z\partial_r A_t}{2}+\frac{\partial_\theta^2 A_r-\partial_r\partial_\theta A_\theta}{2r^2}+\frac{(1-z^3)\partial_z^2-3\partial_z}{2}A_r,
 \end{equation}
 \begin{equation}
 \partial_t\partial_z A_\theta=\Im(\psi_1^*\partial_\theta\psi_1+\psi_2^*\partial_\theta\psi_2)-A_\theta(|\psi_1|^2+|\psi_2|^2)
 +\frac{\partial_r\partial_\theta A_r-\partial_r^2 A_\theta+\partial_z\partial_\theta A_t}{2}+\frac{(1-z^3)\partial_z^2-3\partial_z}{2}A_\theta.
 \end{equation}
\end{widetext}

The rotation is introduced by imposing the angular boundary condition as \cite{Domenech}
\begin{equation}
a_\theta=\Omega r^2,
\label{rotation}
\end{equation}
where $\Omega$ is the constant angular velocity of the disk.
The radius of the boundary disk is set as $r=R$.
The Neumann boundary conditions are adopted both at $r=R$ and $r=0$, $\partial_r h_i=0$ where $h_i$ represents all the fields except $a_\theta$.

The EoMs are solved numerically by the Chebyshev spectral method
in the $z,r$ direction, while  Fourier decomposition is adopted in the $\theta$ direction. The time evolution is simulated by the fourth order Runge-Kutta method. The GPU Computing is used to speed up the calculation.
The initial state at $t=0$ is always chosen to be a homogeneous solution for fixed ($\epsilon,\eta$) when there is no
rotation, which can be obtained by solving the time independent EoMs by  fixing a charge density $\rho$ with the Newton-Raphson method.
Such an initial homogeneous state will evolve by perturbing the system with the rotation.
The confiration after a long time revolution is considered to be stable when the changes of the norm of all fields become smaller than $10^{-5}$ for sufficient long time.
Practically,  a solutions at later time $t$ is used to minus the solutions at $t-\delta t$, when the maximum of the change of the solution become smaller than $10^{-5}$ for sufficient long time (for example $\delta t=100$), then  the solution can be though to be a stable one. We also check that the solution  satisfies the equations of motion with errors less than $10^{-5}$.
Stay in the homogeneous ansartz, there is a critical $\rho_c(\epsilon,\eta)$ above which the two scalar fields will condense with the same value. From numerics we found that $\rho_c(0,0) \sim 4.07$,
then the dimensionless critical temperature is $T_c^0 = \frac{3}{4\pi \sqrt{\rho_c(0,0)}}=0.1183$.
Turning on the direct coupling $\eta$ from $-0.5$ to $0.5$ will not alter the critical temperature,
though a positive/negative $\eta$ will reduce/increase the value of order parameter a bit.
By tuning  $\epsilon$ from $0$ to $0.1$, the critical $\rho_c$ is decreasing, means the
critical temperature is increasing,  for example, the  $\rho_c(0.1,0)=3.7$ indicates a higher $T_c=0.1241$.
In all the dynamic simulation we set $T=0.82 T_c^0$, at which for every combination of $-0.5\leq\eta\leq0.5$ and $0\leq\epsilon\leq0.1$,
the system always have a stable homogeneous solution with the same finite value of order parameters for the two components.

\section{Phase separation in the presence of no rotation}

\begin{figure}[t]
\centering
\includegraphics[trim=2.5cm 10.5cm 5.9cm 9.7cm, clip=true, scale=0.7, angle=0]{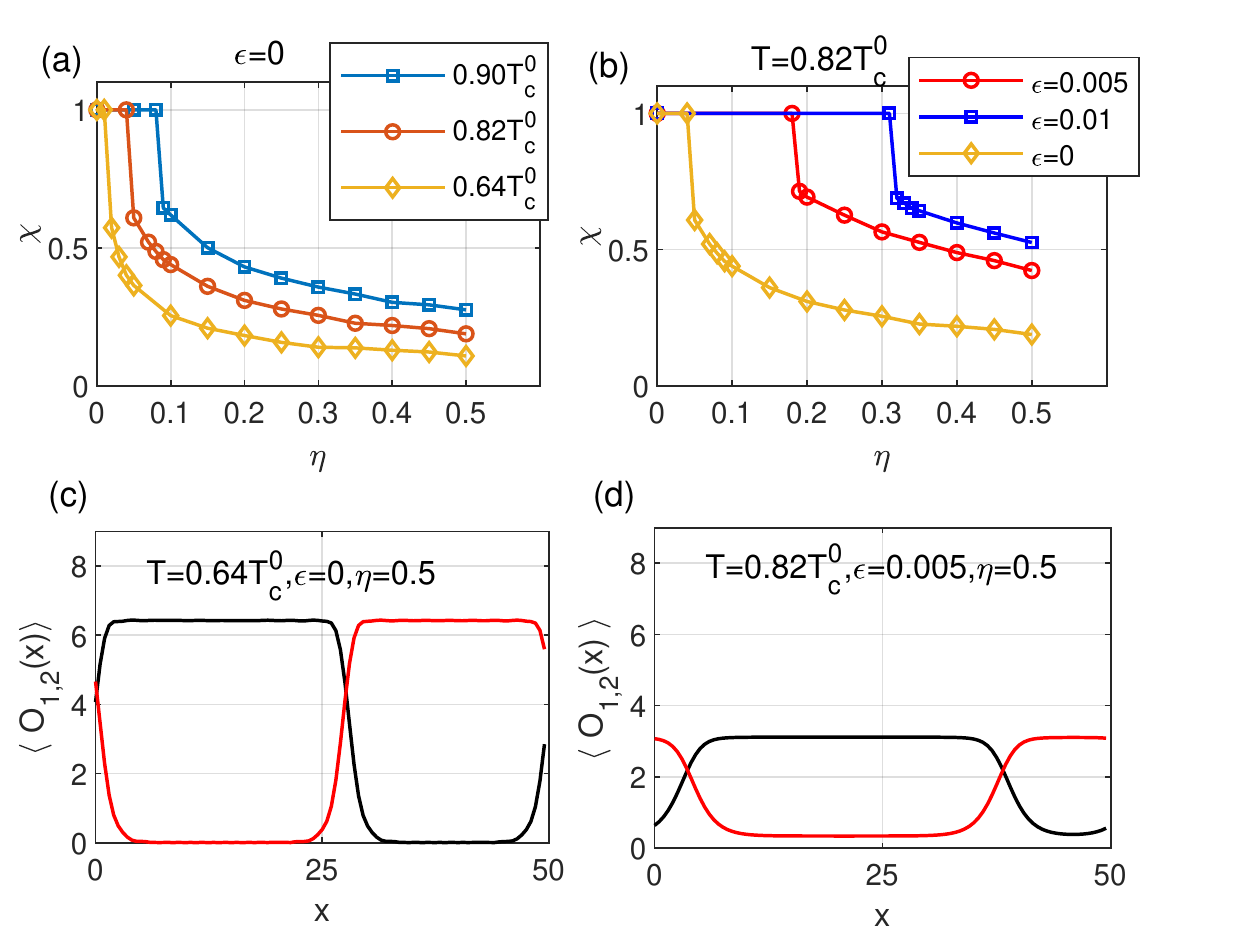}
\caption{(a)-(b) The overlap factor $\chi$ as a function of $\eta$. (c)-(d) The configuration of
 $\langle O_1(x)\rangle$(black line) and $\langle O_2(x) \rangle$(red line) in a phase separated phase.}\label{fig1}
\end{figure}

Phase separation is an inhomogeneous solution that the two condensates
do not overlap spatially. Such an immiscible state can be more stable with a lower free energy than  the miscible
state, which is a result of the repulsive interaction between the
two condensations when the positive $\eta$ is large enough.
The inhomogeneous stable solution can be obtained by solving the full time dependent equation.  At initial time
the system is also in a homogeneous and miscible solution with $\Psi_1=\Psi_2$ at fixed $\epsilon,\eta$, such a homogeneous state might be
a metastable state which can evolve to a final stable inhomogeneous state by perturbing the system with a very small fluctuation, the
real ground state. Without rotation, it is more convinient to adopt the Cartesian  coordinate rather than the polar coordinate  in the
boundary theory, the metric is then $ds^2=\frac{\ell^2}{z^2}\left(-f(z)dt^2-2dtdz + dx^2+ dy^2\right)$. To focus on the phase separation  in the $x$ direction, we take the ansartz that  all the fields are functions of $t,z,x$.  To quantitatively describe the
spatial overlap of the two condensations, we define an integral
\begin{equation}
\chi=\int_0^L dx \frac{|\langle O_1(x) \rangle||\langle O_2(x) \rangle|}{|\langle O_h \rangle|.^2},
\end{equation}
where $\langle O_h \rangle$ is the corresponding homogeneous order parameter at $t=0$, the length $L=50$.

If $\chi= 1$, system shows no phase separation, and it stays in the initial homogeneous state. Otherwise,
if $\chi \ll 1$, it would
be fair to say the system shows phase separation. In the
intermediate case, the system is partially phase-separated
and partially phase-mixed. In Fig. 1(a), we show the overlap integral $\chi$ as a function
of $\eta$ at three different temperatures when $\epsilon=0$. The increase of $\chi$ along temperature
when $\eta=0.5$ indicates that the system is harder to enter the phase separation phase, agrees with the
increased correlation length of both condensations when $T$ is approaching $T_c$. In Figs. 1(b), the $\chi(\eta)$
is shown for three different $\epsilon$ at a fixed temperature, a small $\epsilon=0.01$ will increase the $\chi(\eta)$,
indicates that the two condensations prefers to overlap spatially by increasing the Josephson coupling $\epsilon$.
We also checked that in the case $\epsilon=0.1$, the system is always in the phase overlap state even in the
strong repulsive interaction when $\eta=0.5$, $\chi$ is always to be one.
Two samples of $\langle O_{1,2}\rangle$
of a separated phase with large repulsive interaction are illustrated in Fig.1 (c), and Fig.1 (d). Such an exotic immiscible state has been
experimentally observed in a two-\cite{Hall} and a three-\cite{Stenger} component quantum fluid. Interesting, the phase separated state can also be obtained from a holographic first order phase transition in inhomogeneous black holes\cite{Janik,Attems}.

\begin{figure}[t]
\centering
\includegraphics[trim=3.6cm 10.6cm 3cm 11cm, clip=true, scale=0.53, angle=0]{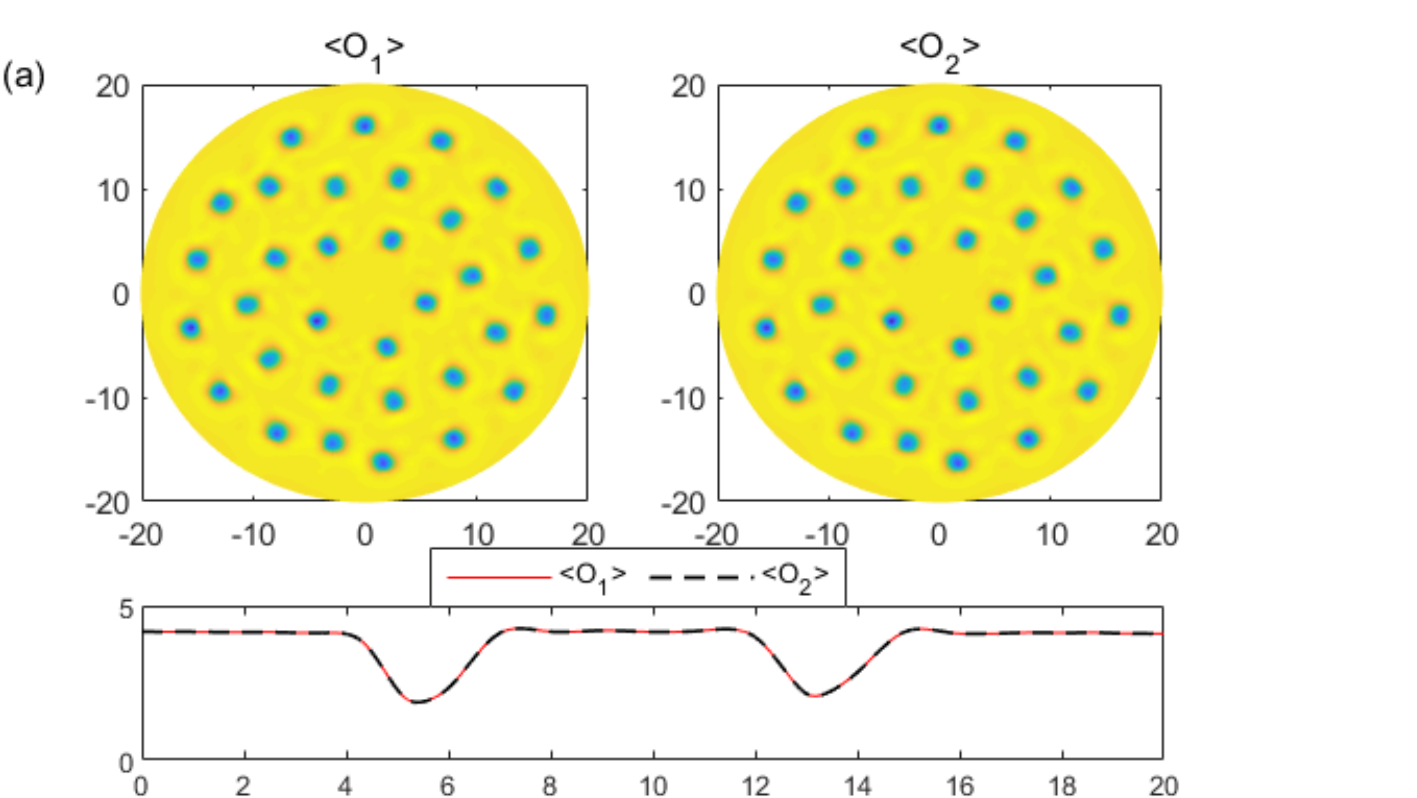}
\includegraphics[trim=3.6cm 10.6cm 3cm 10.5cm, clip=true, scale=0.53, angle=0]{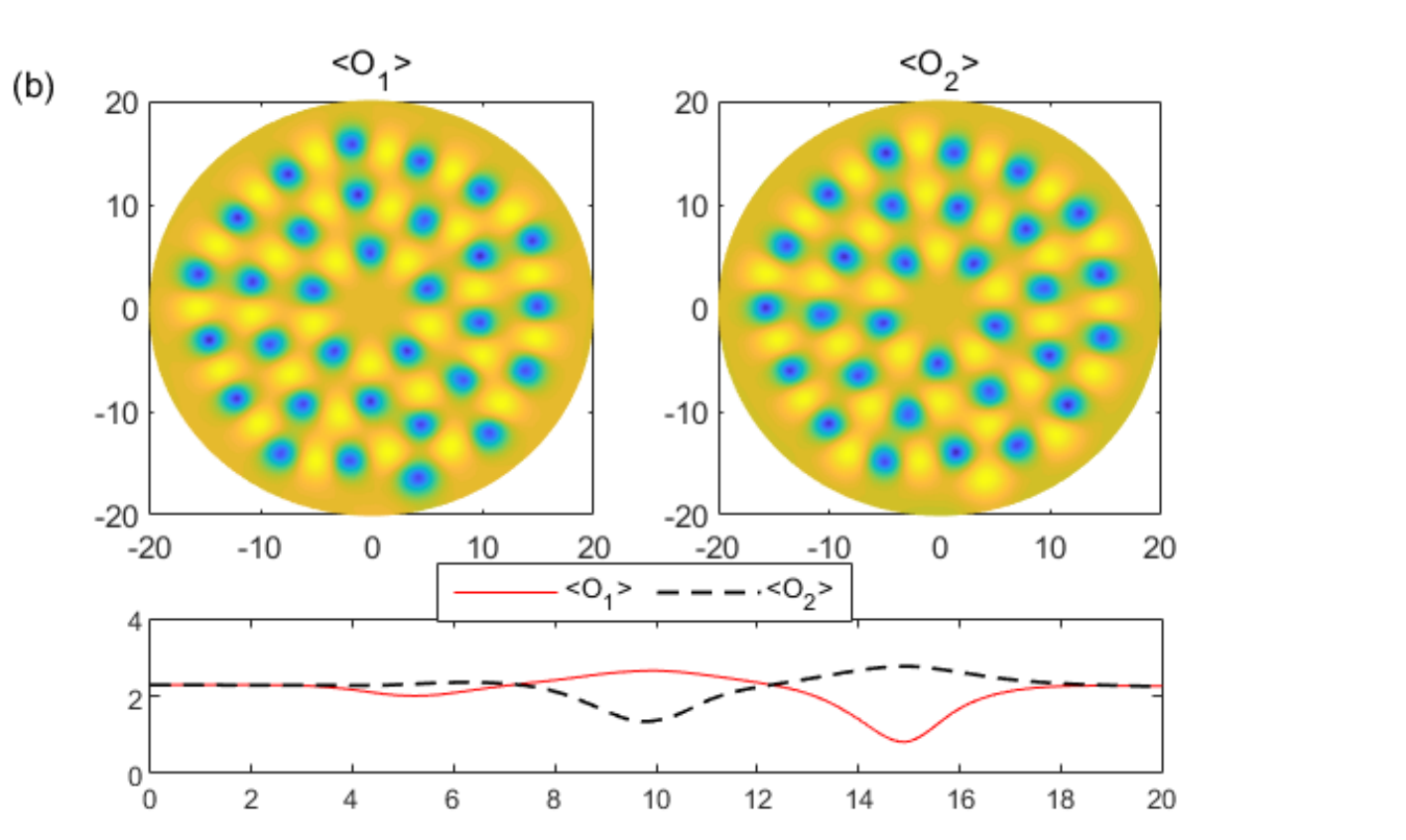}
\includegraphics[trim=3.3cm 11cm 3cm 10.7cm, clip=true, scale=0.53, angle=0]{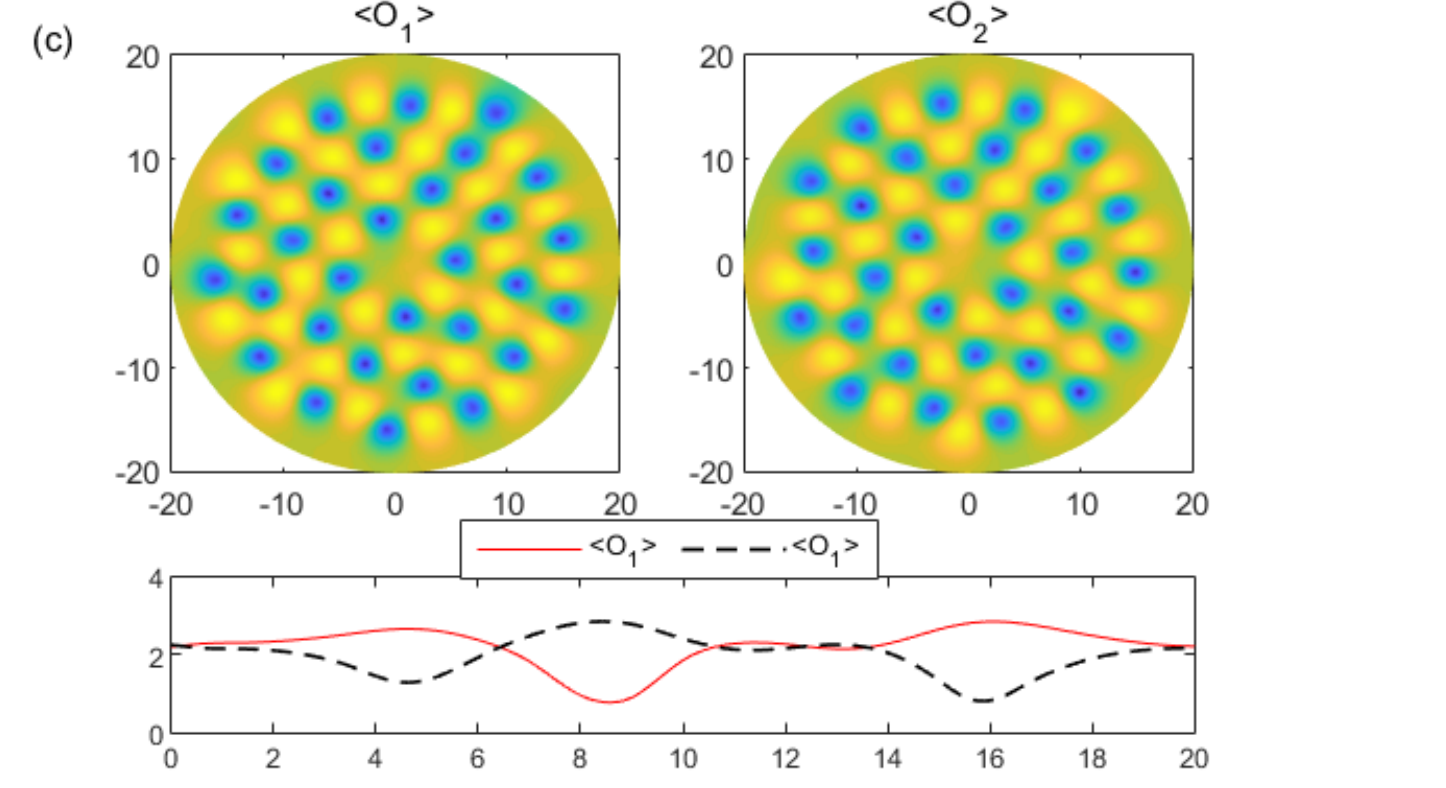}
\includegraphics[trim=4.1cm 10cm 3cm 10.7cm, clip=true, scale=0.53, angle=0]{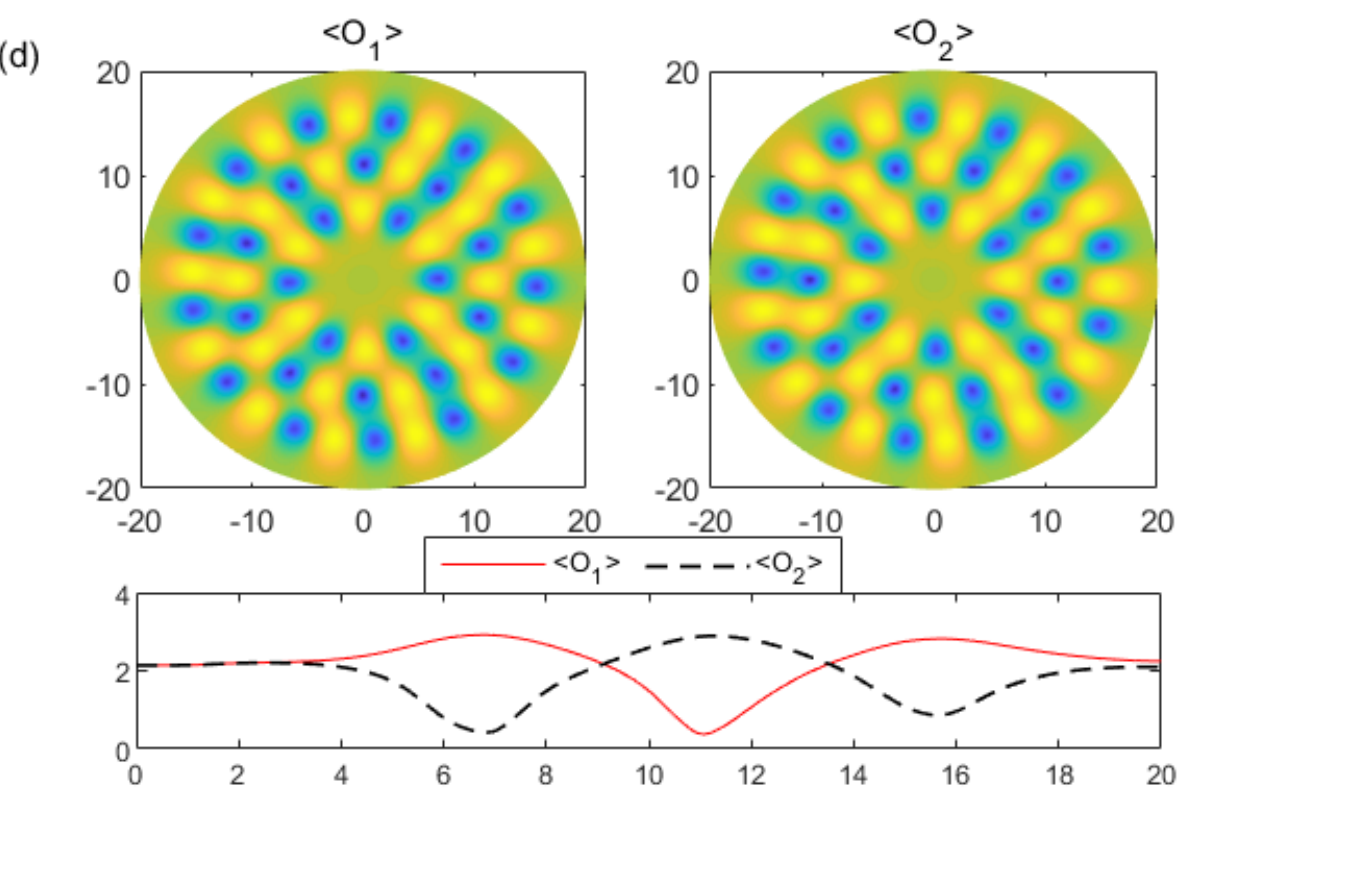}
\caption{Four typical vortex phases at $T=0.82 T_c^0$ and the corresponding radial profiles of $\langle O_{1,2}\rangle$ in the
$\theta=\pi/2$ direction : (a) $\epsilon=0.1,\eta=0.1,\omega=0.1$. (b) $\epsilon=0,\eta=-0.6,\omega=0.1$. (c) $\epsilon=0,\eta=-0.2,\omega=0.1$. (d)$\epsilon=0,\eta=-0.05,\omega=0.1$.}\label{fig2}
\end{figure}

\section{exotic vortex phases under rotation and the phase diagram}

 In the section we move to the case  that the system is in the presence of rotation, focusing on the case when the holographic two component superfluid is at a temperature $T=0.82 T_c^0$. Several kinds of exotic vortex phases have been found.
Firstly, since a finite $\epsilon=0.1$ will prevent the system to enter the phase separation state even $\eta$ is finite, then we expected that the vortices of the two components
coincide, and a trianglar vortex lattice is expected as in the one component case \cite{Xia:2019eje}. A sample result of  $\epsilon=0.1$ and $\eta=0.1$ is shown in Fig.2 (a), even we increase $\eta$ to $0.5$.
the vortices of both components have the same positions，matches the two single vortices solution for the two components locate at the 
same position \cite{Wu}.
Since in the case of  finite $\epsilon$, we will always find  a configuration that the two component vortices take the same position,
 $\epsilon$ is set to  be zero. Then by tuning the inter-component interaction $\eta$ from $-0.6$ to $0.6$, the system
 will demonstrate several exotic vortex phases.
In Fig. 2 (b),(c) and (d), by increasing $\eta$ from $-0.6$ we see triangular lattices ($-0.6\leq\Omega\leq-0.45$), square lattices ($-0.45\leq\Omega\leq-0.1$) and vortex stripes ($-0.1<\Omega<0.05$).
The square lattice is stable, presumably due to the fact that each vortex in one component can
have all its nearest-neighbor vortices to be in the other component\cite{Schweikhard}.These various structures are similar to that obtained by the component G-P equations \cite{Kasamatsu}.

\begin{figure}[h]
\centering
\includegraphics[trim=2.8cm 9.2cm 2.9cm 9.5cm, clip=true, scale=0.5, angle=0]{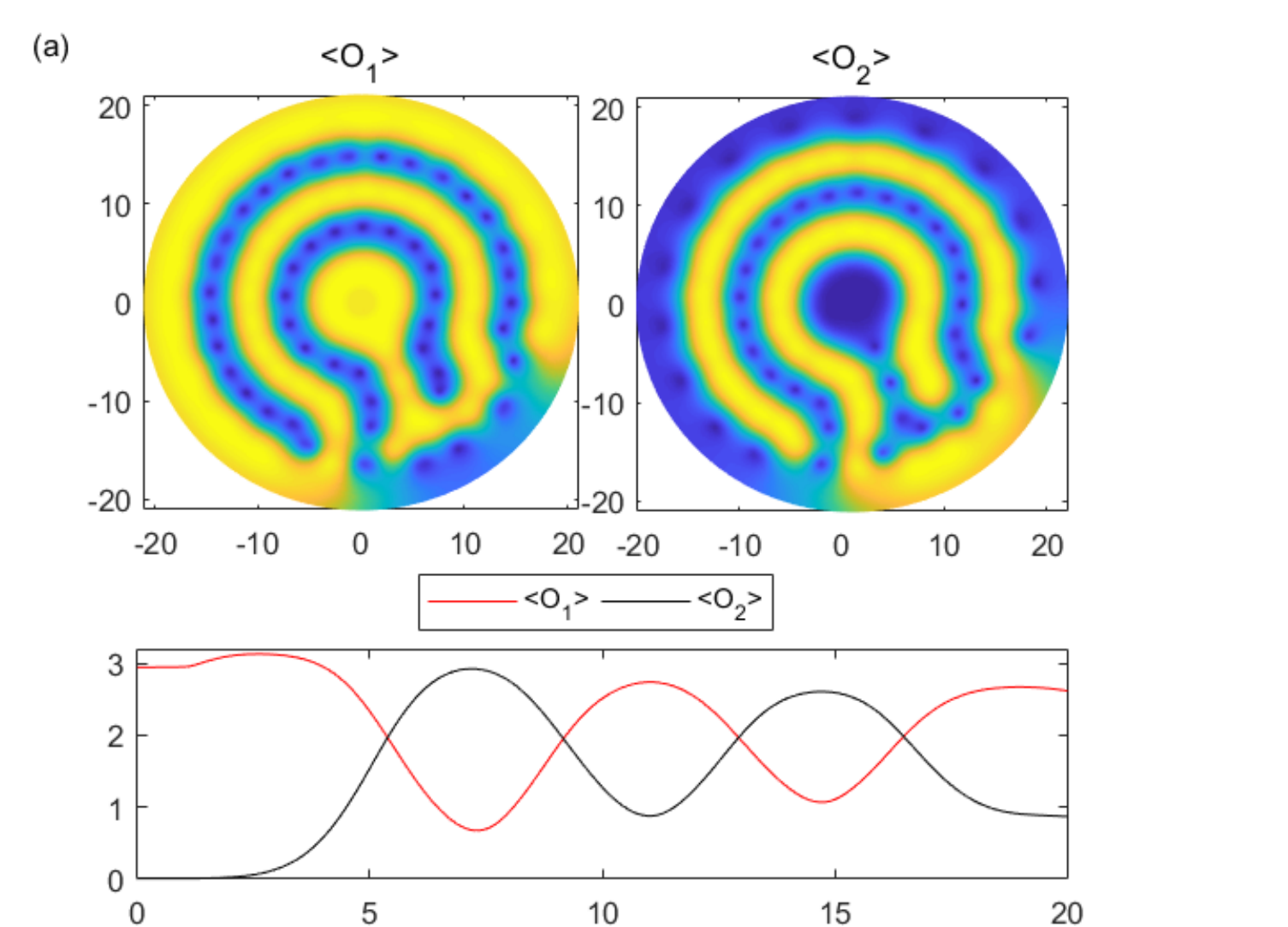}
\includegraphics[trim=2.1cm 9cm 2.9cm 9.5cm, clip=true, scale=0.5, angle=0]{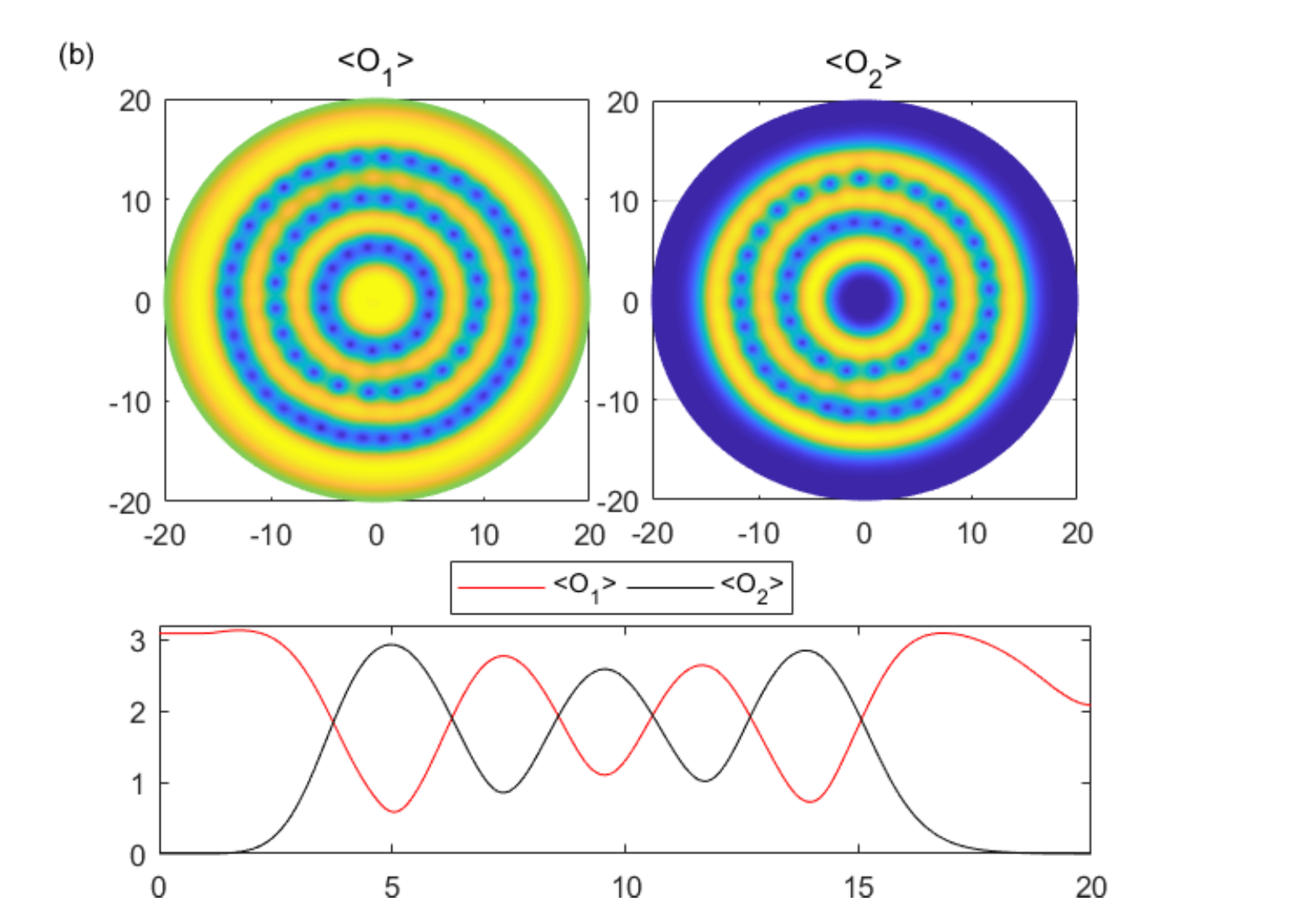}
\caption{Two typical perfect vortex sheet solutions and the corresponding radial profiles of $\langle O_{1,2}\rangle$
 averaged in the $\theta$ direction at $T=0.82T_c^0$: (a) $\epsilon=0,\eta=0.2,\omega=0.15$. (b) $\epsilon=0,\eta=0.2,\omega=0.3$.  }\label{fig3}
\end{figure}
Keep increasing the repulsive interaction to the phase separated region, the vortex sheet is found as plotted in  Fig.3.
In a classical turbulence, the vortex sheet is a thin interface across which the tangential
component of the flow velocity is discontinuous.
In quantum fluid, Landau and Lifshitz firstly proposed the vortex sheets scenario in rotating superfluid \cite{Landau}, almost
at the same time when Feynman published his paper on quantized vortices in superfluid. A quantum vortex sheet solution is that
the vorticity concentrated in line with the irrotational
circulating flow stay between them. A typical picture of vortex sheet can be found in Fig.1 of a review  paper \cite{Volovik},
where the vortices concentrated in circles with a uniform distance between the circles.
However, as a novel quantum state, the vortex sheet had never not been observed
in superfluid $^4$ He due to the  unstable tangential discontinuity against the breakup of sheet into pieces. vortex sheet has been observed in chiral superfluid $^3$ He-A since it can be stable in  due to the confinement of the vorticity within the
topologically stable solitons\cite{Parts}, which may has many physical applications.
The other candidate to demonstrate vortex sheet is a quantum fluid with repulsive two order parameters.
Since the phase separation naturally provides a region of vanishing order parameters of one component while the same region is filled
by the other component, under rotation, the nucleated vortices
merge to form a winding sheet structure like “serpentine” in the order parameter vanishing region instead of forming a periodic lattice.
As shown in Fig.3, in the phase separation region $\eta>0.05$ (see the line corresponding to $0.82T_c^0$ in Fig.1(a).),
we find the exotic vortex sheet solutions. The vortices
of the $\langle O_1\rangle$ are located at the region of the domains of  $\langle O_2\rangle$ component. This can be understood
from the fact that the condensate  of one component works as a pinning potential for the vortices in the
other component due to the phase separation nature \cite{Kasamatsu}. By forming vortex sheets, the condensate achieves remarkable phase separation compared to
a lattice.
Furthermore, the vortex sheets nearly uniformly fill the disk, and the distance $d$ between the layers are equal.

\begin{figure}[t]
\centering
\includegraphics[trim=1.9cm 9.8cm 1cm 9.5cm, clip=true, scale=0.8, angle=0]{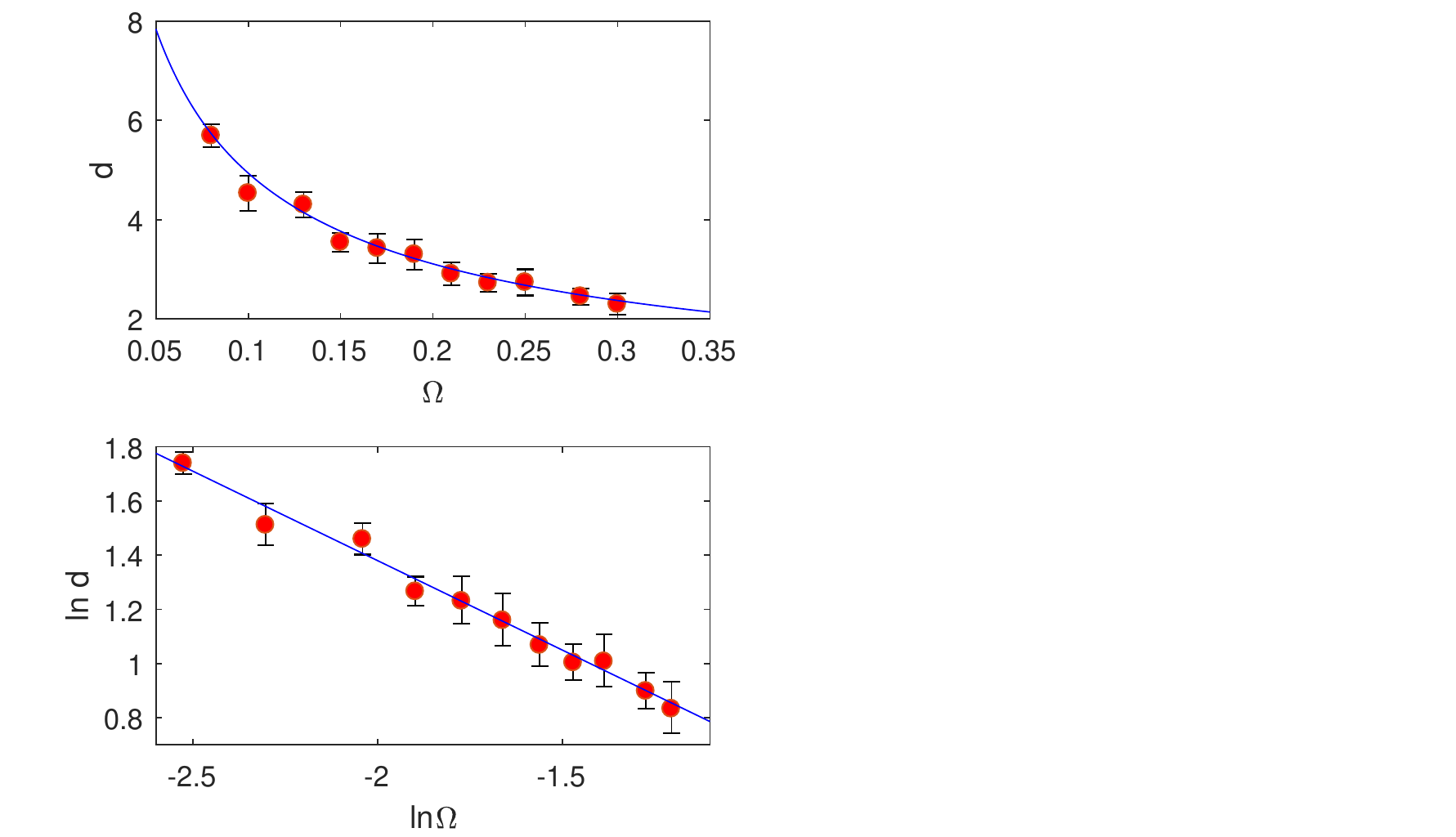}
\caption{(Left):The sheet distance $d$ as a function of $\Omega$ on a $\Omega- d$ plot: $d = \frac{1}{0.95}\Omega^{-\frac{2}{3}}$. (Right):The sheet distance $d$ as a function of $\Omega$ on a Log-Log plot. The straight line is the best fit to $\ln d = a\ln\Omega+b$ with $a=-0.655\pm0.027$ and $b=0.060\pm0.048$.}\label{fig4}
\end{figure}

According to the calculation by Landau and Lifshitz \cite{Landau}, the distance d between sheets is determined by the surface tension $\sigma$ of the soliton and the kinetic energy of the counterflow $(v_n-v_s)$ outside the sheet, where $v_n$ is the normal fluid velocity and $v_s$ is the vortex-free superfluid velocity. In unit volume, the counterflow energy is $\frac{1}{d}\int\frac{1}{2}\rho_s(v_n-v_s)^2dy = \frac{\rho_s\Omega^2d^2}{6}$, and the surface energy is $\frac{\sigma}{d}$,
where $\rho_s$ is the superfluid mass density.
By the minimization of energy, one obtains
\begin{equation}
d = (\frac{3\sigma}{\rho_s\Omega^2})^\frac{1}{3}.
\end{equation}
We confirm that the formula also hold in the two component holographic superfluid, a sample result when $\epsilon=0,\eta=0.2$ is plotted in In Fig.4.

As another important properties of superfluid, the Feynman linear relation between
the excited vortex numbers and angular velocity in one component superfluid may can
naturally generalizes to a two component superfluid as
\begin{equation}
N_j=\frac{M_j\Omega}{\pi \hbar}.
\end{equation}
In the holographic model we also investigated the validity of the Feynman relation
in the two-component system and found no obvious deviations from it,
since the two components are of the same mass then $N_1=N_2$. A sample result is given in Fig. 5 for  $\epsilon=0,\eta=0.2$.

\begin{figure}[t]
\centering
\includegraphics[trim=4.3cm 14.5cm 1.5cm 9.4cm, clip=true, scale=0.85, angle=0]{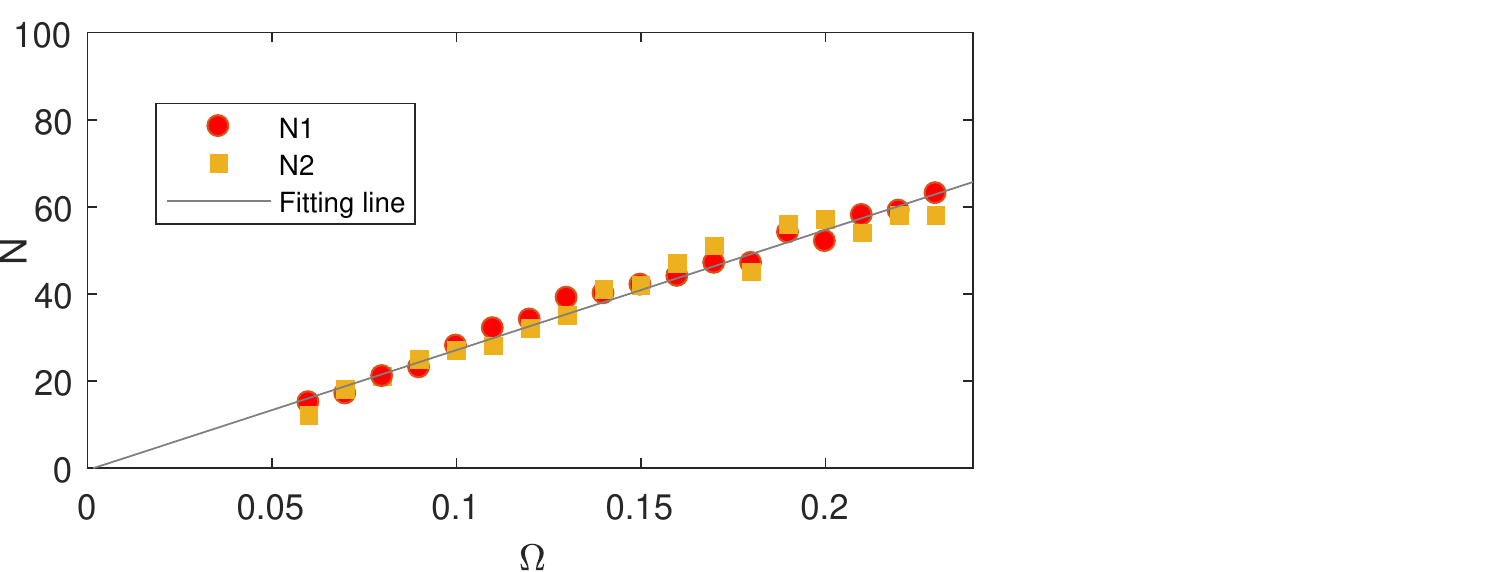}
\caption{Vortex number of both components vs. angular velocity $\Omega$. The straight line is the best fit to $N=a\Omega+b$ with $a=286.62\pm9.77$ and $b=-1.69\pm1.39$. The temperature is fixed to be $T=0.82T_c^0$.}\label{fig5}
\end{figure}

\begin{figure}[h]
\centering
\includegraphics[trim=2.8cm 19cm 1.5cm 2.5cm, clip=true, scale=0.65, angle=0]{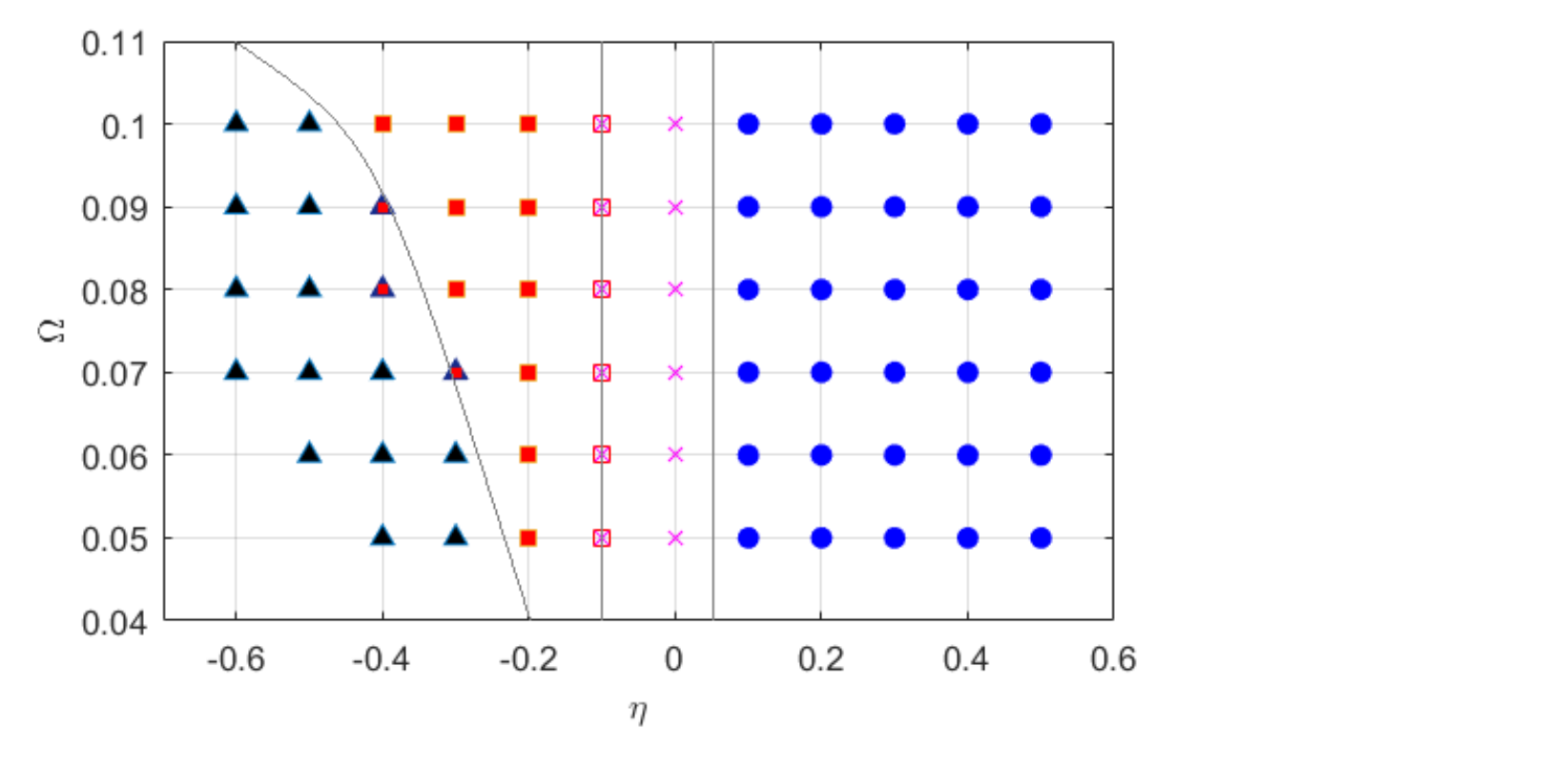}
\caption{$\Omega-\eta$ phase diagram of the vortex states, where $\triangle$ symbolize triangular lattice, $\square$ symbolize square lattice, $\times$ stripe vortex and $\circ$ vortex sheet lattice.}\label{fig6}
\end{figure}

Furthermore,  the phase diagram of the vortex structures in the
intercomponent direct coupling $\eta$ versus rotation-frequency $\Omega$ was investigated, see  Fig.6. The upper limit of the rotation frequency is set by $\Omega=0.1$ while the bottom limit is set by $\Omega=0.05$. The diagram show a transition from triangular lattices to square lattices and then stripe and sheet. When $\eta>0.05$, where is the region of phase separation, it always presents a sheet solution, which confirms our  conjecture that the sheet solution found in Fig.3 under rotation is a result of phase separation.

\section{Conclusions and discussions}
The properties of a two component superfluid obtained from AdS/CFT correspondence in our simulation can be compared to
experiments, for example, in ref \cite{Schweikhard}, the interlaced square lattice similar to Fig.2(c) was observed.
Also in the experiment, the vortex core size of the interlocked  are bigger than the one of
single component, which we can also observe in Fig. \ref{fig2}.
The sheet solution is expected in the highly separated region, which
may can be observed experimentally  in a two component BECs with tunable intercomponent
interactions, which can be deeply in a phase separate region\cite{Thalhammer,Papp}.
The two species are of the same mass, then the vortex core size are the same as shown in
Fig. \ref{fig2} and Fig.\ref{fig3}. Using of different  masses for the two components, we  will realize a coexistence system
of vortices with different vortex-core sizes, then the lattice structure shown in this work may be
changed, which deserves to be studied in future.

Finally we comment on the  properties of the finite
temperature holographic two-component superfluid  that are different from that of the weakly coupled zero temperature
two-component BECs studied in \cite{Ueda,Kasamatsu}. Firstly, in the finite temperature strongly coupled holographic superfluid, the increased correlation length will delay the appearance of the phase separation phase at a larger repulsive coupling $\eta$ (see Fig.\ref{fig1} (a)).
Secondly, the triangle and square lattices (see Fig.\ref{fig2} (a-d)) are less perfect
in the holographic model compared to the perfect lattices founded in a very low temperature rotating spinor BECs\cite{Schweikhard}; the distortions of  regular lattices found here is very likely due to the  relatively high temperature close to $T_c^0$, since we have confirmed that  at higher temperature $T=0.98T_c^0$, the vortex lattice becomes more disorder with a lower translational symmetry, this is also the case we observed in  a single component superfluid for different temperatures\cite{Xia:2019eje}. This is probably due to the fact that the larger vortices at
higher temperature are more closer then the interaction between vortices is large which prevents the
lattices to organized as a perfect lattice.
Employ a gravity dual theory at zero temperature defined in AdS soliton\cite{Nishioka:2009zj}, the vortex lattices with
perfect hexagonal and square symmetry might be  expected to be obtained in single and two-component superfluid respectively.
Thirdly, even at finite temperature, the perfect sheet solutions with accurately equidistant layers were obtained for the first time from holography (see Fig.\ref{fig3}) in the deeply phase separated regime, this is probably due to the strongly coupling nature of the holographic model.
While in a weakly coupled  zero temperature BECs\cite{Kasamatsu}, the distances $d$ between sheets are
not perfect equal which  is harder to calculate when comparing the numerical results to the Landau-Lifshitz  formula.

\emph{Acknowledgements}.---
We thank Li Li and Muneto Nitta for valuable comments. This work is supported by the National Natural
Science Foundation of China (under Grant No. 11675140, 11705005, 11875095).

W.C.Y. and C.Y.X. contributed equally to this work.


\end{document}